\documentclass[acmtog, authorversion]{acmart}

\usepackage{titlesec}
\titlespacing*{\section}
{0pt}{3ex}{1ex}
\titlespacing*{\subsection}
{0pt}{2ex}{1ex}
\titlespacing*{\subsubsection}
{0pt}{2ex}{1ex}

\usepackage{caption}
\usepackage{tabularx}
\usepackage{tabulary}
\usepackage{booktabs}
\usepackage{threeparttable} 

\usepackage{multicol}
\usepackage{subcaption}

\AtBeginDocument{%
  \providecommand\BibTeX{{%
    \normalfont B\kern-0.5em{\scshape i\kern-0.25em b}\kern-0.8em\TeX}}}

\setcopyright{acmcopyright}
\copyrightyear{2023}
\acmYear{2023}

\begin{document}

\title{Abnormal Trading Detection in the NFT Market}
\subtitle{A Statistical Analysis of the NFT Market Structure}


\author{Mingxiao Song}
\affiliation{%
  \institution{Georgia Institute of Technology}
  \city{Atlanta}
  \country{USA}}
\email{songmingxiao38@gmail.com}

\author{Yunsong Liu}
\affiliation{%
  \institution{Georgia Institute of Technology}
  \city{Atlanta}
  \country{USA}}
\email{liuys0118@gmail.com}

\author{Agam Shah}
\affiliation{%
  \institution{Georgia Institute of Technology}
  \city{Atlanta}
  \country{USA}}
\email{ashah482@gatech.edu}

\author{Sudheer Chava}
\affiliation{%
  \institution{Georgia Institute of Technology}
  \city{Atlanta}
  \country{USA}}
\email{sudheer.chava@scheller.gatech.edu}


\begin{abstract}
  The Non-Fungible-Token (NFT) market has experienced explosive growth in recent years. According to DappRadar, the total transaction volume on OpenSea, the largest NFT marketplace, reached 34.7 billion dollars in February 2023. However, the NFT market is mostly unregulated and there are significant concerns about money laundering, fraud and wash trading. The lack of industry-wide regulations, and the fact that amateur traders and retail investors comprise a significant fraction of the NFT market, make this market particularly vulnerable to fraudulent activities. Therefore it is essential to investigate and highlight the relevant risks involved in NFT trading. In this paper, we attempted to uncover common fraudulent behaviors such as wash trading that could mislead other traders. Using market data, we designed quantitative features from the network, monetary, and temporal perspectives that were fed into K-means clustering unsupervised learning algorithm to sort traders into groups. Lastly, we discussed the clustering results' significance and how regulations can reduce undesired behaviors. Our work can potentially help regulators narrow down their search space for bad actors in the market as well as provide insights for amateur traders to protect themselves from unforeseen frauds.
\end{abstract}




\maketitle

\section{Introduction}
NFTs are unique digital assets secured and authenticated by cryptography technologies. Under the auspices of blockchain's decentralized, immutable, and transparent nature \cite{monrat2019survey}, the creation of NFT renders an efficient and secured virtual trading environment with explosive outgrowth. Up until the end of 2021, the market boomed to over \$15 billion for nearly 2 million active wallets \cite{nonfungible2021year}. Although in 2022 the NFT market experienced a cooling-off period as reported by NonFungible \cite{nonfungible2022}, the crypto industry was thus pushed into a more mature stage \cite{nasdaq2022}. As estimated by Grand View Research, the global non-fungible token market size is expected to reach 211.72 billion USD by 2030 \cite{gvr2022}.

However, the pseudonymous nature of on-chain transactions has also created incentives for fraudulent behavior and scams. A common practice is wash trading: buying and selling the same NFT simultaneously to create the appearance of an inflated price and active demand. According to CryptoSlam, about \$18 billion of the trading volume on the LooksRare NFT marketplace, nearly 95\% of the total activity, can be traced to wash trading-related actions \cite{rahul2022}. Also, in reference to the study conducted by Chainalysis, even though not all wash traders are profitable, only 110 suspicious accounts alone generated around \$8.9 million in profits \cite{chainalysis2022}. Besides, there is also a lack of regulative rules for wash trading in the NFT market. Regulatory institutions such as the US Securities and Exchange Commission (SEC) \cite{sec2021} and the Financial Industry Regulatory Authority (FINRA) \cite{finra2017} have taken strict stances against wash trading in traditional financial markets. But due to the novelty of the NFT market, the anonymity and secure essence of the technology, as well as the debate on whether centralized management should be introduced to the crypto market, few regulations have been enacted to clean up wash trading in the NFT market.   

Research related to NFT fraud detection is still limited as fraudulent actions can’t be simply defined by clear patterns, especially facing sophisticatedly designed endeavors to hide adversarial intentions. In the growing phase of the NFT market, active exploration can serve as the cornerstone for future research. Thus, we propose an innovative approach utilizing data mining and machine learning techniques. This corresponds to our goal which is not to give an exact list of abnormal traders but to provide an overview of behavioral patterns and potential insights for future identification of abnormal practices as well as the NFT market structures. The design of our research includes three steps. Firstly, we performed data collection and exploratory data analysis to find distribution patterns and evidence of anomalies as discussed in section 3. Secondly, we designed three types of features including network, monetary, and temporal features which are explained in section 4.1, inspired by research done in traditional financial markets. We then employed K-Means clustering to learn abstruse decision rules and group wallets with comparable actions together in section 4.2. Finally, the clustering result provides an instructive overview of the NFT market structure in regard to accounts, wash trader percentage, and potential group of wash traders, which will be shown in section 5. To the best of our knowledge, this is the first study on abnormal NFT market pattern recognition through machine learning methods. 

Our research serves two groups of audiences. Firstly, uninformed traders and amateurs benefit from our findings. According to the National Cryptocurrency Enforcement Team \cite{mondaq}, more NFT marketplaces should implement automated monitoring tools that can identify suspicious activities. With our research, more characteristics can be used as crucial keys to identify fraudulent behaviors, transactions, and accounts, thus helping marketplaces deliver transparent and liquid pre-trade and post-trade information to all levels of NFT investors. Secondly, regulatory agencies can reference the potential group of wash traders to conduct validation and carry out purposeful research within each of the user behavioral groups we identified. In summary, our research is meaningful in supporting the identification of fraudulent targets in the NFT market while creating a compelling case for improving security scrutiny.

\section{Literature Review}
Some literature provides evidence for the existence of wash trading. Cho. et al.’s paper found that the LooksRare protocol was in charge of most sales at elevated prices \cite{cho2023non}. Tariq et al. conducted statistical analyses such as Benford’s law and clustering effects, proving the pervasiveness of abnormal prices and automated trades \cite{tariq2022suspicious}.


\textbf{Graph-based Approach} The graph-based approach treats user interactions as a network and then identifies graph clusters, and detects collusive groups. Das et al. detected malicious trading behaviors by identifying strongly connected components in the user interaction network \cite{das2022understanding}. Von et al. examined illicit trades by looking into clusters of wallets with no obvious position changes after sequential transactions \cite{von2022nft}. An innovative graph visualization called the NFTDisk was presented by Wen et al. \cite{wen2023nftdisk}. Graph-based Approach is also useful for detecting wash trades in traditional financial markets. Cao et al. analyzed wash trading patterns using digraphs and dynamic programming \cite{cao2015detecting}. In another research done by Victor et al., the graph-based strongly connected component identification method was again used for tracing wash trades on decentralized crypto exchanges \cite{victor2021detecting}.

\textbf{Statistical Approach} Statistical approaches analyze abnormal behaviors from a quantitative point of view. For example, Serneels introduced three ways to detect wash trading with criterions such as close-loop token trades, closed-loop value trades, and high transaction volumes \cite{serneels2023detecting}. A more mathematically-based procedure was proposed by Pelechrinis et al. who built a regression model for profit prediction and filtered out anomalous transactions that generate above the average profit \cite{pelechrinis2022spotting}. Cong et al. quantified wash trading on cryptocurrency exchanges by exploring first significant-digit distributions, size rounding, and tail distributions of transaction volume \cite{cong2022crypto}.  Pennec et al. predicted ETH/BTC trading volumes using a regression model on web variables and wallet variables \cite{le2021wash}. 

\textbf{Data Mining and Machine Learning Approach} Data mining and machine learning techniques have become increasingly popular for fraud detection in financial markets. Thai et al. employed unsupervised learning methods such as k-means, Mahalanobis distance, and unsupervised SVM to detect anomalous behaviors in the Bitcoin network \cite{pham2016anomaly}. Monamo et al. also tested K-means and trimmed K-means clustering algorithms on currency, network, and average neighborhood features in Bitcoin transactions \cite{monamo2016unsupervised}. Finally, Gubran et al.’s paper provided a comprehensive overview of most state of art fraud detection techniques used in the general financial market \cite{al2021financial}. In summary, these techniques can help identify patterns that may not be visible to the human eye, detect subtle changes in transactions, and uncover previously unknown patterns of fraud that may have gone unnoticed using traditional methods.

Overall, previous work mostly focuses on fraud detection in traditional financial markets or cryptocurrencies. Also, only a limited amount of behavioral patterns for wash trading have been detected in the NFT market and there are more intrinsic characteristics to be exploited. Finally, cutting-edge data mining and machine-learning techniques haven’t been fully uncovered in the field of non-fungible tokens. Its ability to detect complex patterns and help with automation in decision-making processes can be beneficial in the exploratory process for abnormality detection in the NFT market.

\section{Data}
In this section, we discuss data collection and preliminary data analysis which serve as the foundation for the quantitative definition of different behavioral patterns.

\subsection{Data Collection}
Before introducing the dataset we used, we clarify some terms used in Non-Fungible Token trades. NFT Tokens are one-of-a-kind digital assets verified on the blockchain. NFT Collections refer to a collection of unique NFT tokens that are issued by the same artist. Some popular collections in the Art category include CryptoKitties, Azuki, and CryptoPunks \cite{nadini2021mapping}. The NFT Collections can be traded in NFT marketplaces, which are hubs for buying and selling NFTs. Some popular marketplaces include Opensea, Cryptoslam, and SuperRare \cite{DBLP:journals/corr/abs-2105-07447}. There are also NFT exchanges where orders can be placed by users and matched or executed by the exchange. As part of the transaction cost, a gas fee is required to compensate for the efforts and resources taken by miners to verify and add the transaction to the blockchain. The NFT buyers need a wallet to store and manage their NFT tokens and the wallet is identified by a unique wallet address. An NFT transaction operated by users can either be a sale, when a buyer purchases an NFT from a seller in exchange for cryptocurrency or other assets, or a transfer, when the owner of an NFT sends it to another wallet address without any exchange of payment.

Our research mainly focuses on transactions that occurred on an NFT marketplace, specifically OpenSea. Because the amount and characteristics of wash trading behaviors can be different among collections, we wanted to formulate a representative dataset. Therefore, we first used the Reservoir API to acquire collection-level information for 20,000 collections on the OpenSea marketplace. By eliminating inactive ones by ordering collections based on their tokenCounts and volume, we ended up with 5000 collections worth to be analyzed. From this pool of candidates, we randomly selected 100 collections to be a representative microcosm of the entire market, containing large, medium, and small collections with different frequencies of suspicious activities. We then used the Moralis API to retrieve all transaction histories for NFT tokens in those 100 collections ever since they were minted. For each transaction, the entries in the record include token\_address (uniquely identify an NFT collection), token\_id (uniquely identify an NFT token), from\_address (seller’s wallet address), to\_address (buyer’s wallet address), value (transaction value in Ethereum), block\_number (block number on blockchain for saving the transaction record), and block\_timestamp (transaction timestamp). In total, we gathered over 1 million transactions with 252,924 distinct wallets. In support of our analysis, we also retrieved historical Ethereum to USD and Gas to Ethereum exchange rate data from Yahoo Finance. 

Finally, we conducted data cleaning and removed unnecessary information and noisy data. Data validation was implemented since our dataset consists of data from multiple sources. We also performed data integration and transformation to manage all the data in an SQLite database that can be easily retrieved and analyzed.

\subsection{Exploratory Data Analysis}
To have an overall look at the unusual behaviors and irregularities in the NFT market, we performed a preliminary data analysis utilizing traditional methods of abnormality detection. In this step, we don’t aim to identify abnormal transactions but want to find evidence that shows the existence of abnormal activities. We first examined the first digit distribution of transaction prices against Benford’s law, a common mathematical tool used in detecting fraud. Also, we used a common pattern in trading, round number clustering. Behavioral studies found that people tend to use multiples of 10 in decision-making. Lastly, we inspected the distribution of transaction price, focusing on whether it exhibits fat tails characterized by a power law. 

\begin{figure}[h]
    \centering
    \includegraphics[width=0.7\linewidth]{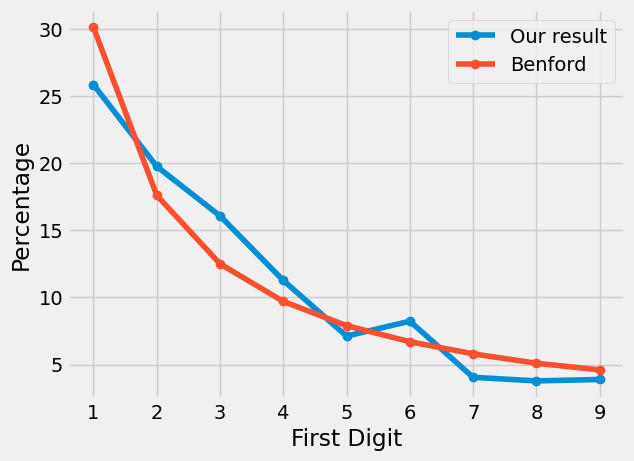}
    \caption{Our observation of the dataset's first digit distribution vs. expectation of the distribution according to Benford's law.}
\end{figure}

Before taking those three approaches, we first removed all transactions with a price of 0 on which the above methods cannot be applied. To explore the distribution of trade prices’ first digit, we grouped transactions by their price’s first significant digit and calculated the percentage of all nine-digit classes. Figure 1 shows that our dataset exhibits a roughly similar exponentially decreasing line as that of Benford’s law. However, to justify whether our dataset follows Benford’s law, we performed a Chi-squared test with the null hypothesis being that the first digits in our dataset follow Benford’s law. With a confidence level of 5\% and a degree of freedom of 9, the Chi-squared test yields a Chi value of 24688, which is significantly larger than the critical value. Therefore, the test rejects our null hypothesis at a confidence level of 5\%. 

Next, we examined the round number clustering effect. We rounded transaction prices to 0 decimals and grouped transactions by their rounded price. By visual examination of the distribution of rounded prices from 0 to 80, we found that the distribution roughly exhibits a clustering at multiples of 5. However, there are abnormally high frequencies at certain rounded price points. For example, the frequency of 37 ETH is over 30,000, more than 5 times higher than the next highest frequency in the range from 0 to 80. 

\section{Model}

To categorize NFT users into groups, we defined the problem as an unsupervised clustering problem with no ground truth. We built representative features and let the model learn to cluster users such that users within each cluster have some convergent behavioral patterns. 

\subsection{Feature Engineering}

Based on the results found in exploratory data analysis, we built patterns that have both statistical and practical significance and can reflect the motivation behind wash trading behaviors. By surveying existing literature that uses machine learning techniques for fraud detection \cite{monamo2016unsupervised, pham2016anomaly, bhattacharyya2011data}, we categorized features into three categories: network features, monetary features, and temporal features. 

\begin{table*}[t]
\begin{threeparttable}[b]
\centering
\begin{tabular}{@{}llp{11.5cm}@{}}
\toprule
Category         & Feature                                      & Description      \\ \midrule
Network          & In-Degree                                    & Total number of sellers the wallet has interacted with as a buyer \\
                 & Out-Degree                                   & Total number of buyers the wallet has interacted with as a seller \\
                 & Unique In-Degree Ratio                       & Total number of distinct sellers the wallet has interacted with divided by in-degree \\
                 & Unique Out-Degree Ratio                      & Total number of distinct buyers the wallet has interacted with divided by out-degree \\
                 \hline
Monetary         & Total In-transaction Volume                  & Total volume of transactions with the wallet as a buyer in USD \\
                 & Total Out-transaction Volume                 & Total volume of transactions with the wallet as a seller in USD \\
                 & Average In-transaction Volume                & Average amount bought in USD \\
                 & Average Out-transaction Volume               & Average amount sold in USD \\
                 & SD of In-transaction Volume                    & Standard deviation of volume bought in USD \\
                 & SD of Out-transaction Volume                   & Standard deviation of volume sold in USD \\
                 & Profit from Transfers                        & Profit gained from transfers \\
                 & Profit Ratio                                 & Sell price minus buy price divided by buy price for all non-zero valued sells \\
                 & Transfer Ratio                               & Number of transfers a wallet made divided by the total number of transactions \\
                 & Relative-Sell                                & Average ratio of selling price to collection EMA7\tnote{1} selling price \\
                 & Relative-Buy                                 & Average ratio of buying price to collection EMA7 buying price \\
                 \hline
Temporal         & In-transaction Interval                      & Average time interval in terms of days between each in-transaction \\
                 & Out-transaction Interval                     & Average time interval in terms of days between each out-transaction \\
                 & Diff Time Interval                           & Difference between interval in-transaction and interval out-transaction for a wallet \\
                 & Max Trans                                    & Maximum number of transactions per day \\
                 & Avg Trans                                    & Average number of transactions per day \\
                 & SD Trans                                    & Standard deviation of the number of transactions per day \\
                 & Avg Minted Days                              & Average number of days since an item being minted for each transaction for each wallet \\
                 & Market-Trend Buy                             & Activeness of a wallet during bear and bull markets\tnote{2} in regard to buying events \\
                 & Market-Trend Sell                            & Activeness of a wallet during bear and bull markets in regard to selling events \\
                 & Buy-ATR                                      & Activeness of a wallet during volatile versus stable markets\tnote{3} in regard to buying events \\
                 & Sell-ATR                                     & Activeness of a wallet during volatile versus stable markets in regard to selling events \\ \bottomrule
\end{tabular}
\begin{tablenotes}
\item [1] 7-day exponential moving average reflects heat in the market in the short term.  
\item [2] The state of a market is determined using 7-day exponential moving average of transaction price. 
\item [3] Volatility of a market is measured by the average true range (ATR).
\end{tablenotes}
\end{threeparttable}
\caption[]{Feature Description.}
\end{table*}

Network features aim to reflect abnormalities in account interactions. Transaction records could be modeled as a directed multi-graph with wallets as nodes and transactions as edges. The directions of the edges follow the flow of the transaction, distinguishing the accounts as buyers and sellers. We observed that most accounts in the NFT marketplace make few transactions with a large group of different accounts. And accounts that make a high number of transactions with a small set of other accounts are more likely to be wash traders due to the fact that wash traders profit from actively participating in the market and trading with the same accounts repeatedly. With such observation, the degree of connections and unique connections an account has served as the perfect tool to capture an account's interactions in the market.

Monetary features are designed to capture abnormalities reflected in transaction sizes in USD. The absolute value, the average, and the standard deviation of the monetary volume construct a comprehensive view of the flow of capital. The profit actually gained by the trader is also an important factor. As mentioned in the introduction, though not all wash traders are profitable, excessive gain is still a crucial identification of wash trading behaviors. Finally, we described users' actions relative to market trends and quantified whether there is an obvious difference in behavioral patterns during bear and bull terms of the collections. 

Lastly, temporal features could summarize a wallet's trading habits from the time dimension. Like other financial markets, compared with retail investors in the market, special groups of traders may trade with high-frequency, regularity, activeness, and other characteristics, which can be reflected by the temporal features we designed.

Table 1 details and describes all the features within each category. These features listed are wallet-based, meaning each of the 252,924 wallets has all features calculated. We standardized by transforming the features such that they have a mean of zero and a standard deviation of one. After designing these features, we calculated the correlation within each of the three feature types. This helps us eliminate some highly correlated features we initially designed with the identification threshold of 0.9.




\subsection{K-Means Clustering}
Without ground truth given, we wanted the model to identify potential grouping patterns and understand the underlying structure of the data, making our task an unsupervised clustering problem. The choice of the clustering algorithm depends on the requirement of the problem, the size of the data as well as the relative performance of each model in practice. Compared to several common clustering models such as Density-based clustering and Hierarchical clustering, K-Means clustering makes the most suitable and explainable predictions for our dataset while performing fast computing with our high dimensional dataset.

K-means cluster data into K distinct groups by iteratively partitioning the data into clusters and adjusting the centroid of each cluster until they converge to stable positions \cite{hartigan1979algorithm}. In our example, each wallet is represented by a multi-dimensional vector with each dimension being a feature calculated above. This produces a set of m points $(x_1,....,x_m)$ in which $x_i \in R^n$, where $n=26$.

\subsubsection{Selection of K}  

A key problem with K-Means clustering is how to find the number K. With the goal of analyzing the customer composition in the NFT market, the number of clusters should be explainable such that it is not producing too many partitions while the number is not too small so that no valuable separation is provided. We started with two graphical techniques to determine the optimal number of clusters.

Firstly, the elbow method works by plotting the within-cluster sum of squares (WCSS) against the number of clusters and identifying the elbow point on the plot. The number at the elbow point is then the one that produces the optimal outcome as adding more clusters doesn’t reduce WCSS significantly. Figure 2 shows the visualization. Though the elbow is not as obvious, we could still use a knee locator to find the elbow point at number 7. It also reflects that the number of clusters within a reasonable range doesn’t affect the performance so much. Next, we calculated the Davies-Bouldin index (DBI) which aims to maximize the similarity within clusters and minimize the similarity between clusters. The lower the value, the better the clustering result. As shown in Figure 2, the global minimum occurs at around 13 and 14. However, considering explainability, dividing the dataset into 13 groups was too much when we need to define a trader type for each group. Thus, we considered the other local minimum which occurs around number 8. Finally, we used the silhouette coefficient to select the optimal number of clusters. It measures how similar an object is to data points within its own cluster compared to other clusters. Higher values indicate better clustering results since it indicates a strong separation between clusters. In order to further filter the candidates provided by the previous two methods, we used the silhouette coefficient to evaluate candidates 7, 8, and 9. As shown by our calculation, these three ways of partitioning have corresponding coefficients of 0.3, 0.29, and 0.28.

\begin{figure}[h]
    \centering
    \includegraphics[width=0.7\linewidth]{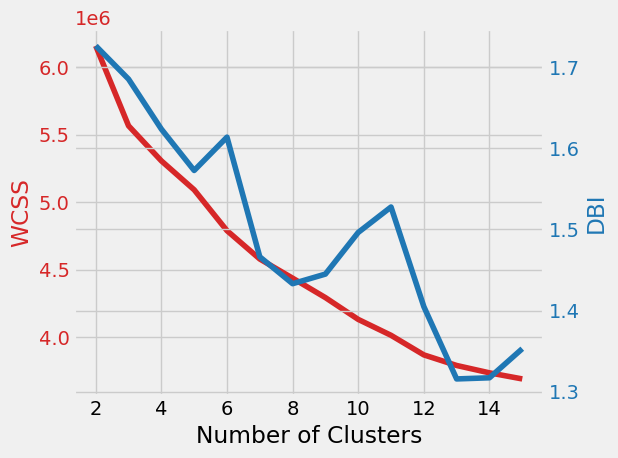}
    \caption{WCSS and DBI vs Number of Cluster Plot.}
\end{figure}

In summary, combining the results from these three methods with the fact that these three divisions all generate groups containing similar sets of data points, we chose k to be 7. Theoretically, it performs well enough divisions while avoiding delicate splits in the general market.

\section{Results}

With the lack of ground-truth structural analysis of the NFT market, we used machine learning evaluation techniques to prove the stability and validity of the algorithm, and then through visual, statistical, and cluster analysis, we assigned interpretable labels to each user group.

\subsection{Result Validation}

To assess the performance of the model, we first performed cross-validation. We shuffled split the dataset into 10 samples, run the trained K-Means model on each of these subsets, and compare two clustering evaluation indices, Sum of Square Error (SSE) and DB-index. We observed that the model performs stably with 10 subsamples and that the train and test portion of each set generates clusters with similar qualities as well. No overfitting problem is presented with our model.


\subsection{Visual Analysis}

To visualize such a high-dimensional dataset, we employed Principal Component Analysis (PCA) to find features that explain the largest variance. Setting the number of principal components to two for simpler visualization, we found the cumulative explained variation to be 25.66\%. Figure 3 visually distributes all data points according to the two principal components. Class 0 in color green locates in the upper left with a high value of both PC1 and PC2. Class 3 in the color black tends to have a high value of PC1 and Class 2 in the color light blue tends to have a high value of PC2. Class 1 is the majority class in red with low values in both PC1 and PC2. Although the clusters are not clearly separated, it could be due to the mixed nature of NFT traders, which makes it hard to set distinguishable bounds. Thus, what is more important is to assign meaningful labels to these clusters by looking into the statistics in the next section.

\begin{figure}[h]
    \centering
    \includegraphics[width=0.75\linewidth]{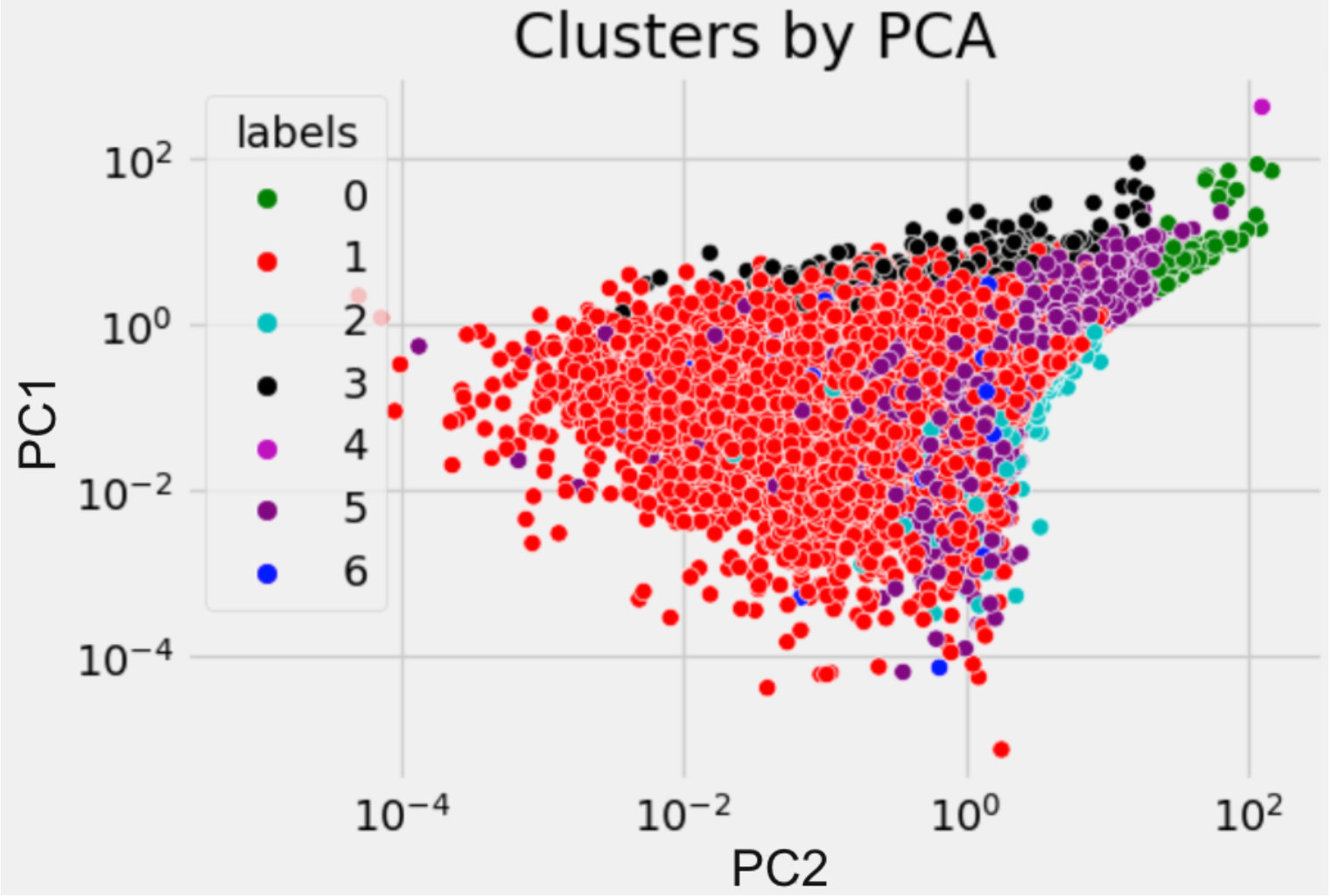}
    \caption{PCA Visualization of the Clustering Result.}
\end{figure}

\subsection{Statistical Analysis}
In this section, we compare feature statistics of the seven clusters in order to find obvious quantitative characteristics that can define particular groups. Among the seven clusters, there are two small clusters, two medium ones, and three relatively large clusters. We then performed statistical comparisons among clusters. Given the large number of features we have as well as multiple clusters, we used the Exploratory Platform to help us with data analysis and visualization. We first constructed radar maps for each type of feature to reflect the general relative characteristics of these clusters, as the mean value of the features for each class is reflected in the map. Take the temporal feature radar map as an example, Class 4 is highly skewed and thus greatly stretches the map so we excluded Class 4 and compared the rest of these classes on relatively the same scales. Now Class 3 has a large mean in features related to the number of daily transactions, while Class 2 stands out in interval in-transaction and diff time intervals. Class 5 also yields lower than average diff time interval and above the average interval in/out-transaction.


\begin{figure}[h]
    \includegraphics[width=\linewidth]{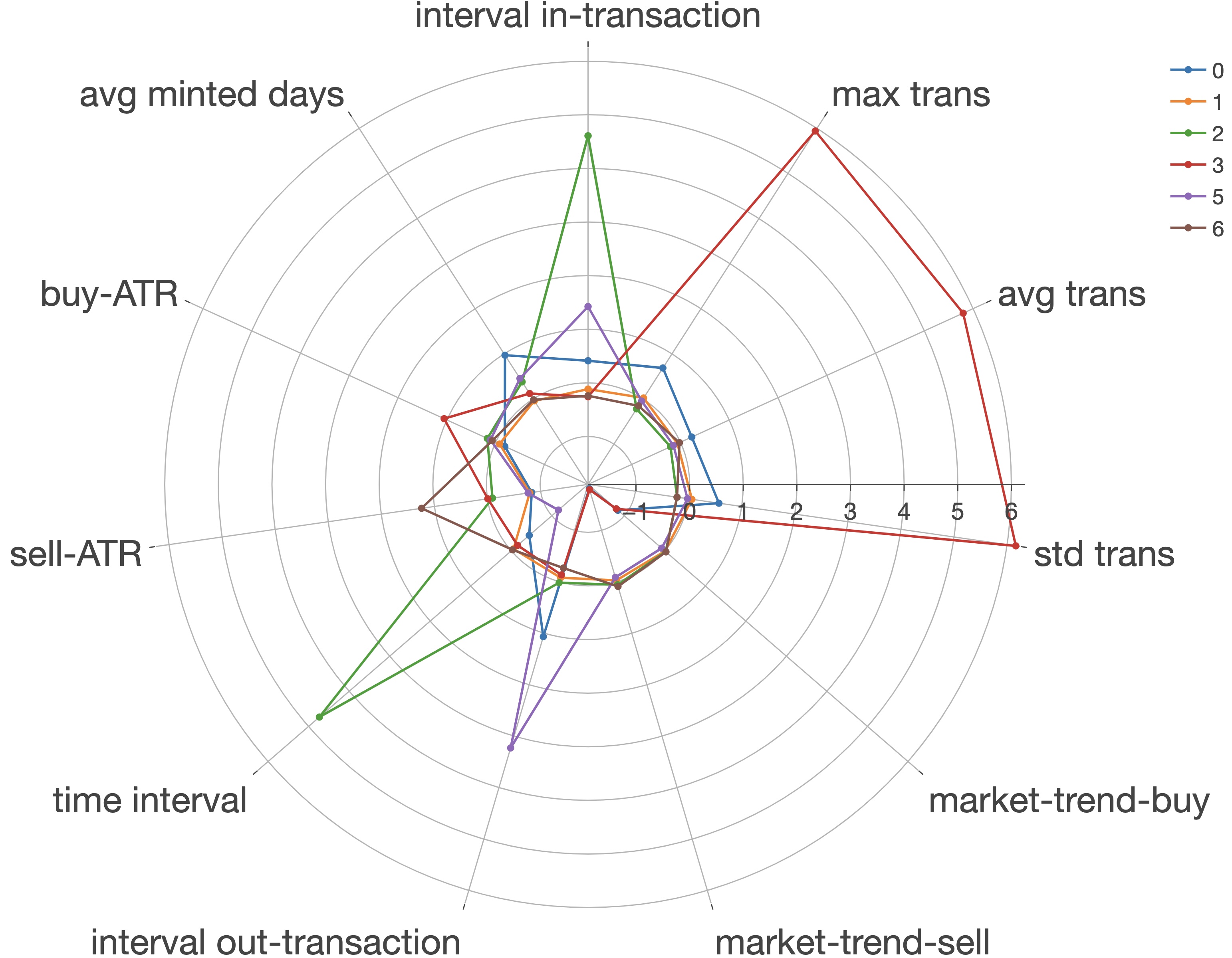}
    \caption{Radar Map for Temporal Features.}
\end{figure}





Besides the radar graphs, we also analyzed boxplots for individual features, with which, instead of simply comparing the mean, we took variance and outliers into account as well. These statistical observations we made provide clues for forming definitive labels for all clusters, explaining user behavioral patterns with jargon in the NFT market, and identifying potential wash trader groups in the next section.

\subsection{Cluster Analysis}
Apart from the aforementioned analyses, we also turned them into explainable facts understandable by actual participants in the NFT market. Upon visual and statistical analysis, we then summarized unique properties for each class and identified them to be some specific types of traders in the market.

\textbf{General Market: Class 1 (137,623 accounts)}

Class 1 with 137,623 accounts represents the market majority and serves as the benchmark since it contains no skewed features.

\textbf{Hodler: Class 6 (96,420 accounts)}

The 96,420 wallet addresses in Class 6 are identified as holders, which is a jargon describing traders buying and holding the asset for a long time. This group of traders essentially invest, never sell, and believe in the asset’s long-term value. From our analysis, the most direct evidence is that all accounts have zero out degree, meaning that the selling activity is never conducted. Another fact is that they tend to invest right after the asset is initially minted.

\textbf{Inactive Accounts: Class 2 (5,145 accounts)}

Class 2 is labeled as inactive users. Their trading patterns are summarized by large intervals between buy activities and buy-sell activities, which means that this group of users doesn’t frequently buy and upon buying doesn’t trade owned products as often.

\textbf{Institutional Accounts: Class 4 (7 accounts)}

Institutional accounts manage a larger amount of money and assets compared to general retail investors. This cluster of accounts is clearly identified in the early phase of analysis as they contain outstanding values in many features and no matter what cluster number we input into the K-Means model, these seven accounts are always aggregated with similarity. These seven accounts generate all the outlier values for in-degree, out-degree, unique-in, and unique-out degrees, as well as the count of transfer activities. Given the limited number of samples in this set, we investigated individual account activities on Etherscan, a block analytics platform for Ethereum. Among these seven accounts, there are institutional accounts for NFT Collections such as AI Cabone, Doodles, and Space Doodles, and for NFT marketplaces such as Gem Swap, and Nimbus.

\textbf{Collectors: Class 0 (122 accounts)}

This relatively small group of accounts is identified as collectors, who establish values and ideas through their purchases and influence the whole NFT market. Different from speculators, investors who enter the market just to be profitable and make money, collectors look for NFT collections with a unique narrative and value the artistic value of the asset. Based on our analysis, they have properties such as high buy and sell volume, relatively large intervals between buy and sell activities, and buying when the collection is quite mature. They also make high profits by setting the selling price higher than the collection average. These all comply with the general understanding of collectors as they need time to evaluate the asset with aggregated insights and analysis.

\textbf{Wash Traders: Class 3 and Class 5 (13,607 accounts)}

Finally, two classes are left as candidates for wash traders. They are Class 3 and Class 5 resulting in 13,607 addresses. No clear labels can be given to these traders but they have some potential characteristics that match our understanding of wash trading behaviors. We found that entities from Class 3 tend to have high in and out-degree, meaning that they actively trade but likely with the same set of accounts. This can be explained by the fact that wash traders always trade with the same set of subaccounts either to inflate price or trading volume. They also have a high transfer ratio and a large number of transactions per day, validating the previous finding that most of their actions are transfers among subaccounts. Finally, they tend to buy and sell when the market is relatively stable since a volatile market is more risky and thus harder to profit as wash traders. On the other hand Class 5 with 11,359 wallets tend to buy and sell in high volume and have large intervals between consecutive in and out transactions. They tend to buy when a collection is relatively mature, meaning that the collection has been minted for a long time. These properties are identifiable features for wash traders.

\section{Conclusion}

Based on the seven clusters we derived from K-Means, five of them are marked with specific labels while the other two are potential wash trader candidates, concluding that the wash trader percentage in the NFT market is 5.38\%. Our analysis is helpful for regulatory agencies since we provide a structural analysis of the whole NFT market, and based on the quantitative features we designed, regulators can only focus on a subset of the general market when constructing supervision rules. Open-source platforms can also be built to provide our findings to the general public and help them make comprehensive judgments. In summary, the research method we have taken can be used as an innovative inspiration that more researchers can use to conduct in-depth analysis.

There are still some limitations of our work that can be addressed in future studies. Firstly, the lack of ground truth brings uncertainty and bias upon evaluation. Thus, future research could either resort to authoritative resources, similar studies, or any open-source dataset to validate our results or provide more detailed divisions. Secondly, jumping out of the scope of identifying wash trading accounts, another potential improvement is to detect wash trading transactions, which better reflect the flow of abnormal dynamics. Finally, since we only took wash traders as our anomaly detection target, more scandalous activities such as money laundering, pump-and-dump, and rug-pull, can be analyzed in similar ways.

With the conclusion of this paper, we hope to attract more people who are interested in protecting the NFT market order and increasing the credibility and stability of the crypto market, with the goal of formalizing a more sophisticated and mature marketplace.

\clearpage

\section*{Acknowledgement}
We gratefully thank Andrew Hornback for his helpful advice during our research process. We sincerely thank peers from the Georgia Tech FinTech Lab for their extensive support. We also thank participants of the Web3 ATL Conference and the Georgia Tech UROP Symposium for their invaluable feedback.

\bibliographystyle{ACM-Reference-Format}
\bibliography{nft-reference}

\clearpage

\appendix

\twocolumn[{
\centering
\includegraphics[width=0.88\linewidth]{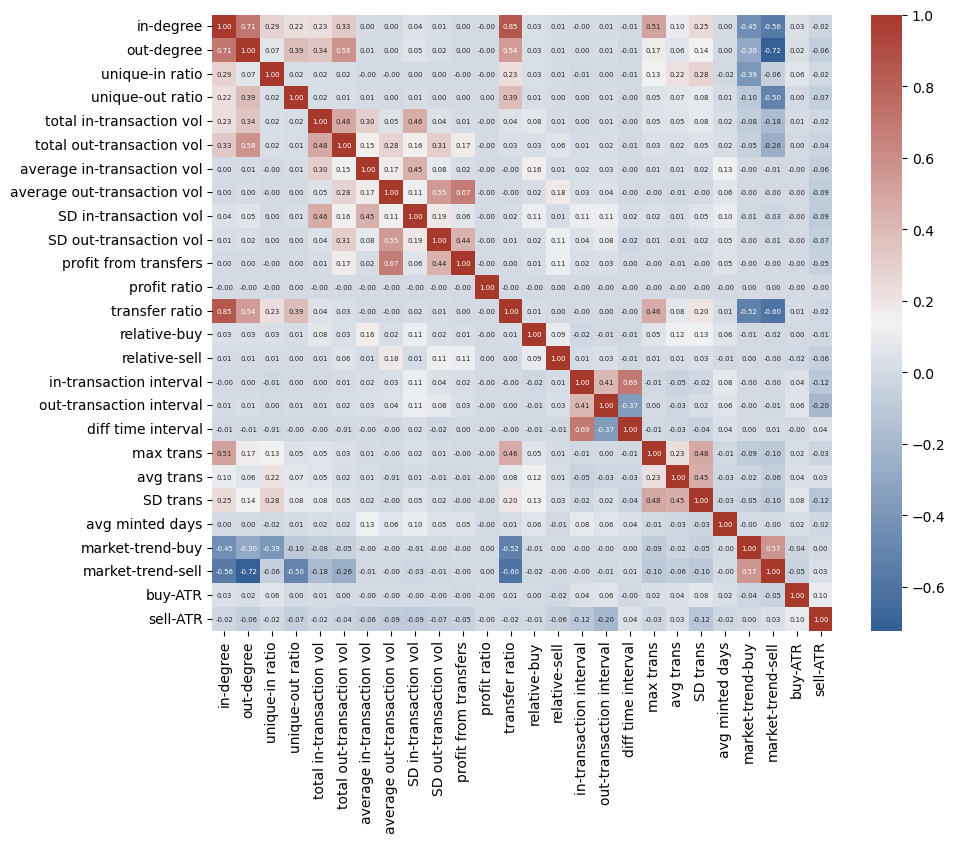}
}]

\newpage

\section{Correlation}

In support of the reproduction of the feature engineering step, we provide the correlation matrix in the figure on top. Highly correlated features within the same category are eliminated. The correct implementation of the aforementioned features with our dataset should resemble the above benchmark. 

\section{SSE and DB-Index for Result Validation}

For the reproduction of the whole training process, we provide the cross-validation results for SSE and DB-Index. These results show that clustering consistently exhibits similar performance with the training set and the test set over 10 iterations.

\section{Data and Code}

For reproduction purposes, data\footnote{\url{https://drive.google.com/file/d/1rejw_dr5snvdsik3zylgYWdJOKpYOZxp/view}} and code\footnote{\url{https://github.com/erikLiu18/NFT-Abnormality-Detection}} are provided. 


\begin{figure}
     \centering
     \begin{subfigure}[h]{0.5\textwidth}
         \centering
         \includegraphics[width=0.5\textwidth]{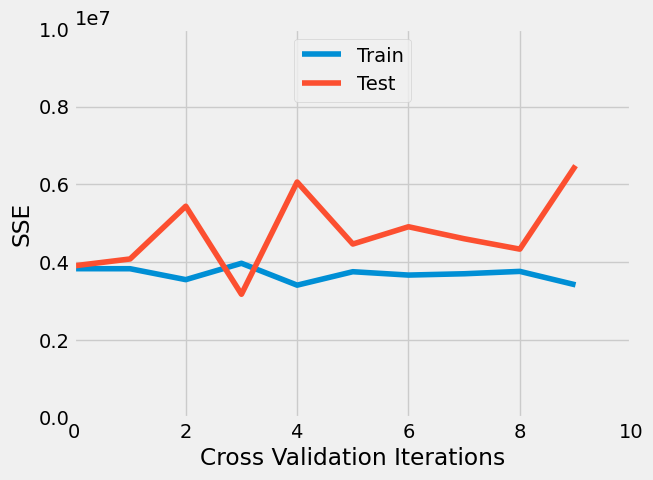}
         \caption{Cross Validation SSE.}
     \end{subfigure}
     \par\bigskip
     \begin{subfigure}[h]{0.5\textwidth}
         \centering
         \includegraphics[width=0.5\textwidth]{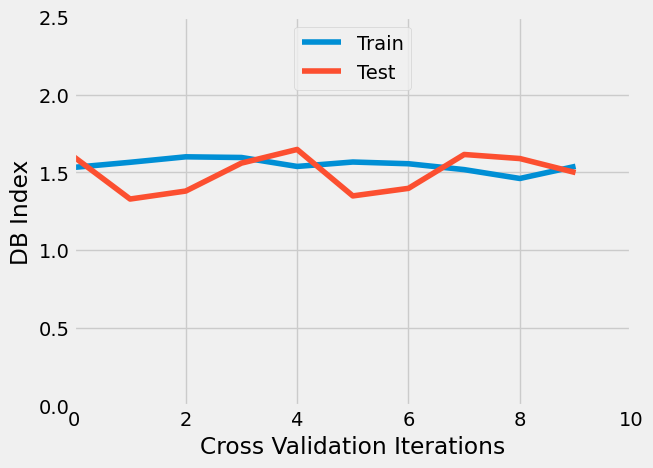}
         \caption{Cross Validation DB-Index.}
     \end{subfigure}
\end{figure}

\end{document}